\documentclass[aps,showpacs,twocolumn,prb]{revtex4-1}
\usepackage{hyperref}
\usepackage{graphicx}
\usepackage{amsmath,amssymb,amsfonts}

\begin{document}

\title{Tuning the Kondo effect in Yb(Fe$_{1-x}$Co$_{x}$)$_{2}$Zn$_{20}$}
\author{Tai Kong} 
\author{Valentin Taufour}
\author{Sergey L. Bud'ko}  
\author{Paul C. Canfield}
\affiliation{Ames Laboratory, U.S. DOE, and Department of Physics and Astronomy, Iowa State University, Ames, Iowa 50011, USA}

\begin{abstract}

We study the evolution of the Kondo effect in heavy fermion compounds, Yb(Fe$_{1-x}$Co$_{x}$)$_{2}$Zn$_{20}$ (0$\leqslant$ x $\leqslant$ 1), by means of temperature-dependent electric resistivity and specific heat. The ground state of YbFe$_2$Zn$_{20}$ can be well described by a Kondo model with degeneracy $N$ = 8 and a $T_K\sim$30 K. In the presence of a very similar total CEF splitting with YbFe$_2$Zn$_{20}$, the ground state of YbCo$_2$Zn$_{20}$ is close to a Kondo state with degeneracy $N$ = 2 and a much lower $T_K\sim$ 2 K. Upon Co substitution, the coherence temperature of YbFe$_2$Zn$_{20}$ is suppressed, accompanied by an emerging Schottky-like feature in specific heat associated with the thermal depopulation of CEF levels upon cooling. For 0.4$\lesssim$ x $\lesssim$ 0.9, the ground state remains roughly the same which can be qualitatively understood by Kondo effect in the presence of CEF splitting. There is no clear indication of Kondo coherence observable in resistivity within this substitution range down to 500 mK. The coherence re-appears at around x$\gtrsim$ 0.9 and the coherence temperature increases with higher Co concentration levels.

\end{abstract}
\maketitle

\section{Introduction}

The $RTM_{2}$Zn$_{20}$ ($R$ = rare earth elements. $TM$ = transition metals) series of compounds were discovered two decades ago, in 1997\cite{Nasch97}. They crystallize in a cubic, CeCr$_2$Al$_{20}$ type structure (space group Fd$\bar{\text{3}}$m) where $R$ ions occupy a single crystallographic site. The nearest, and next-nearest neighbors of $R$ ions are all Zn and thus varying the transition metal does not significantly alter the local environment of the $R^{3+}$. Even though more than 85$\%$ of the atomic constituents are zinc, these compounds exhibit myriad physical properties depending on the rare earth and transition metal that are involved\cite{shuang2007,Jia07,Jia09,Canfield08,milton}. When the rare earth element is Yb, there are six closely related Yb$TM_{2}$Zn$_{20}$ heavy fermion compounds for $TM$ = Fe, Ru, Os, Co, Rh, Ir\cite{milton,Mun12}. Among these six compounds, YbCo$_{2}$Zn$_{20}$ has the largest Sommerfeld coefficient, $\gamma\sim$7900 mJ/mol-K$^{2}$. This value is comparable to the record holding YbBiPt\cite{YbBiPt}, and is more than an order of magnitude larger than the other members in this family, which have $\gamma$ values ranging from 520 mJ/mol K$^2$ for YbFe$_2$Zn$_{20}$ to 740 mJ/mol K$^2$ for YbRh$_2$Zn$_{20}$\cite{milton}. The reason behind the dramatic difference between YbCo$_2$Zn$_{20}$ and the other five Yb$TM_2$Zn$_{20}$ compounds is still not clear, although band structure calculations reveal that the 4f level is closer to the Fermi level in YbCo$_{2}$Zn$_{20}$ than in YbFe$_{2}$Zn$_{20}$. Upon Co substitution, the d band is gradually filled, which is accompanied by a drop in d band energy\cite{Tanaka10}.

Being the heaviest of the Yb$TM_{2}$Zn$_{20}$ compounds, YbCo$_{2}$Zn$_{20}$ has been studied intensively ever since its discovery\cite{milton}. Upon application of pressure, there is an indication that the heavy Fermi liquid regime can be suppressed, followed by the appearance of an antiferromagnetic ordering for $P >$ 1 GPa\cite{Saiga08}. It has, therefore, been argued that YbCo$_{2}$Zn$_{20}$ is close to a quantum critical point (QCP)\cite{Saiga08,Nakanishi10}. Similar suppressions of Fermi liquid regime under pressure were also observed for $TM$ = Fe/Rh/Ir\cite{Kim13,Matsubayashi09,Matsubayashi10}. In the case of YbFe$_{2}$Zn$_{20}$, $\gamma$ = 520 mJ/mol K$^2$, and is thus likely to be further away from a QCP in terms of the Doniach diagram, indeed, a much higher critical pressure, around 10 GPa was proposed in order to reach a QCP\cite{Kim13}. With an effective negative pressure induced by Cd substitution, the hybridization between Yb$^{3+}$ 4f electrons in YbFe$_{2}$Zn$_{20}$ and the conduction electrons becomes weaker\cite{Avila16}. Apart from pressure induced ordering, metamagnetic transitions were reported for YbCo$_{2}$Zn$_{20}$ at high magnetic fields\cite{Shimura11,Takeuchi11b,Shimura12,Honda14}. Crystalline electric field (CEF) schemes for YbCo$_{2}$Zn$_{20}$ have been proposed based on specific heat and anisotropic magnetization measurements. The first and the second excited CEF levels are around 10 K and 25 K above the ground state doublet\cite{Takeuchi11,Romero14}. Inelastic neutron scattering measurements show some excitations that might be related to these proposed CEF schemes\cite{Kaneko12}. Experimentally, band structure had only been reported for YbCo$_{2}$Zn$_{20}$ via quantum oscillations\cite{Ohya10}, with heavy ground state being strongly suppressed by increasing field. In a zero field limit, the mass of the quasiparticles was extrapolated to be 100-500 times the free electron mass. 

Given that (i) the first and second nearest neighbors of Yb$^{3+}$ in Yb$TM_2$Zn$_{20}$ do not have $TM$ and (ii) as $TM$ changes from Fe to Co the values of $T_K$ and $\gamma$ change by an order of magnitude, it is of interest to see how the strongly correlated electron state evolves in Yb(Fe$_{1-x}$Co$_x$)$_2$Zn$_{20}$ for 0$\leqslant$x$\leqslant$1. In this paper, we report the temperature-dependent resistivity and specific heat data on 19 members of the Yb(Fe$_{1-x}$Co$_x$)$_2$Zn$_{20}$ series and track the effects of band filling and disorder on the coherence and Kondo temperatures as well as amounts of entropy removed by thermal depopulation of CEF levels versus Kondo state.

\section{Experimental Methods}
 
Single crystals were grown using a high temperature solution growth technique\cite{shuang2007,milton}. The starting molar stoichiometry was Yb:$TM$:Zn = 2:4:94. Bulk elemental material (Yb from Ames Laboratory Material Preparation Center (99.9$\%$ absolute purity); Fe (99.98$\%$), Co (99.9+$\%$) and Zn (99.999$\%$) from Alfa Aesar) were packed in a frit-disc crucible set\cite{Canfield16} and sealed in a silica tube under $\sim$0.25 bar of Ar atmosphere. The ampoule assembly was then heated up to 900 $^{\circ}$C over three hours; dwelt at 900 $^{\circ}$C for 10 hours and then cooled to 600 $^{\circ}$C over 100 hours. At 600 $^{\circ}$C, the remaining Zn rich solution was decanted from the crystals that formed on cooling. Samples were cut and polished so that magnetic field is applied along the [111] crystallographic direction. Resistivity was measured using a standard 4-probe technique in a Quantum Design (QD) Physical Property Measurement System (PPMS). Epotek-H20E silver epoxy was used to attach Pt wires onto the samples. Specific heat was measured using a QD PPMS. A dilution refrigerator option or a $^{3}$He option was utilized to perform measurements down to 50 mK or 500 mK. Elemental analysis was performed via wavelength dispersive spectroscopy (WDS) using an electron probe microanalyser of a JEOL JXA-8200 electron microprobe.

\section{Experimental Results}

\begin{figure}[tbh!]
\includegraphics[scale = 0.35]{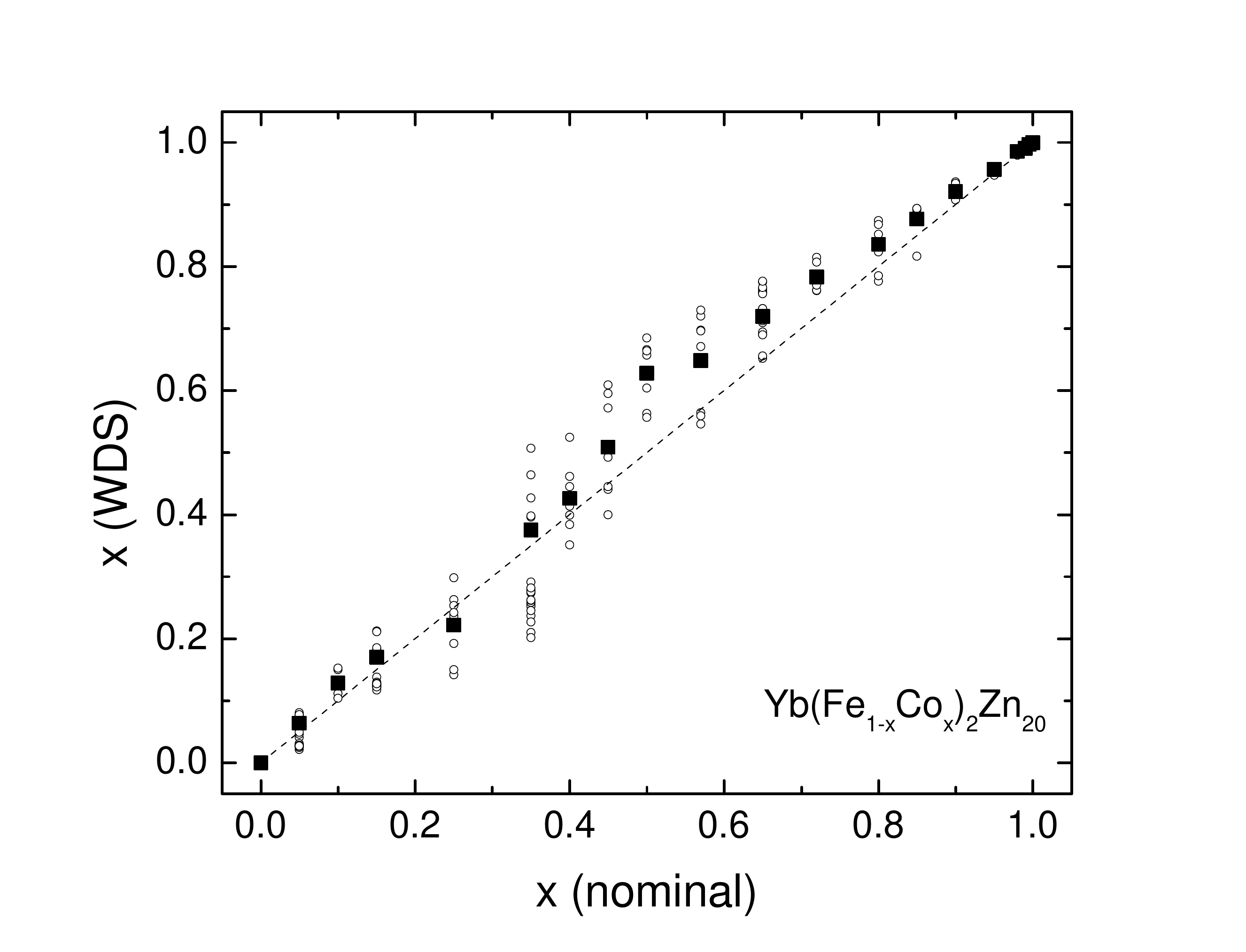}
\caption{Co concentrations determined by WDS as a function of nominal Co concentration values. Solid squares represent averaged WDS values. Hollow circle data points show measured values at different spots of samples. Dashed line is a guide for the eyes with a slope of 1.}
\label{WDS}
\end{figure}

Fig.~\ref{WDS} shows the WDS determined Co concentration as a function of nominal Co concentration. Although the average value of the WDS determined concentration changes close to linearly with the nominal concentration, in the middle of the substitution range, the variation of the substitution level is large. The variation is based on WDS results measured at different spots on the sample as well as different samples in the same batch. Near the two ends of the series, the variation in substitution level is significantly smaller. 
 
\begin{figure*}[tbh!]
\includegraphics[scale = 0.65]{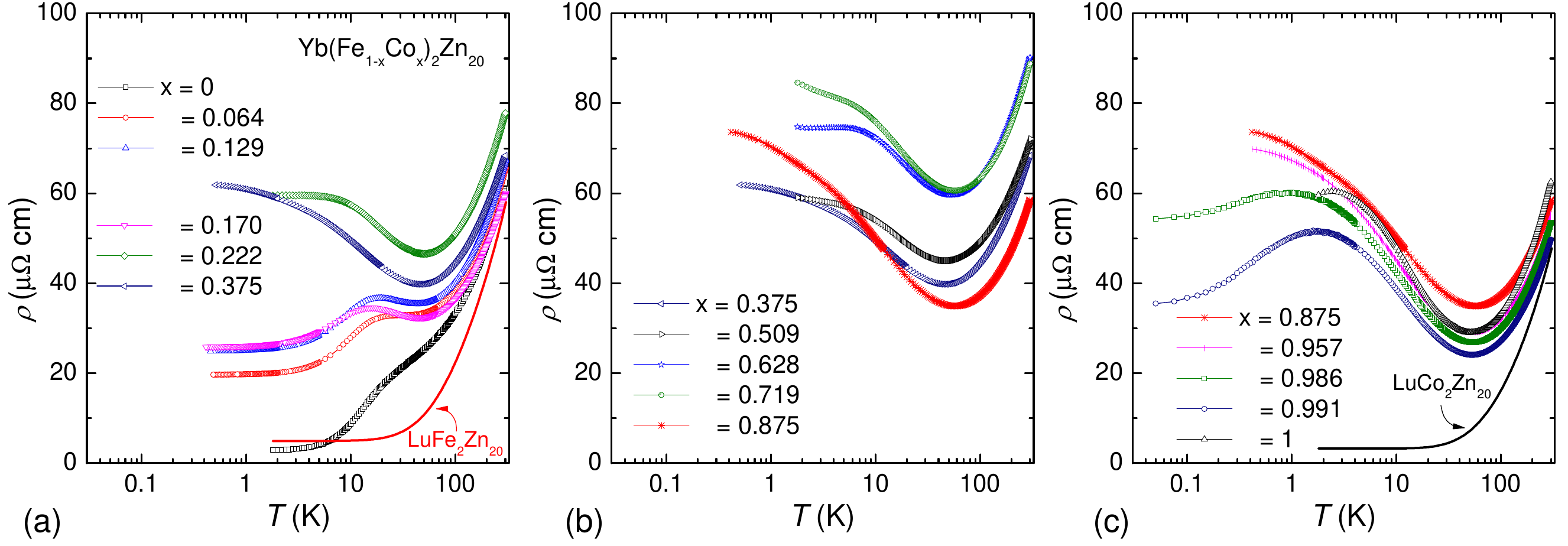}
\caption{(Color online) Zero-field temperature-dependent resistivity of Yb(Fe$_{1-x}$Co$_x$)$_2$Zn$_{20}$ for a selective of Co concentrations on semi-log plots. Resistivity of non-magnetic LuFe$_2$Zn$_{20}$ and LuCo$_2$Zn$_{20}$ are also shown for comparison.}
\label{R}
\end{figure*}

Fig.~\ref{R} shows the temperature-dependent resistivity data of Yb(Fe$_{1-x}$Co$_x$)$_2$Zn$_{20}$ on semi-log plots. Fig.~\ref{R}(a) shows data closer to the pure YbFe$_2$Zn$_{20}$ side. At x = 0, the resistivity of YbFe$_2$Zn$_{20}$ shows a broad shoulder at $\sim$ 40 K, after which it goes into a Kondo coherence. At low-temperature, YbFe$_2$Zn$_{20}$ manifests a $T^2$ temperature-dependence with the coefficient of $T^2$ resistivity, $A$ = 0.054 $\mu\Omega$ cm/K$^2$[\onlinecite{milton}]. As the Co concentration increases, the resistivity value at 300 K increases, most likely as a result of increasing amount of scattering due to chemical disorder (i.e. Fe/Co). In addition, the temperature at which resistivity starts to decrease, after the low temperature shoulder or local maximum, gradually shifts to lower temperatures, indicating a lowering of the Kondo coherence temperature. At x = 0.064, the low-temperature resistivity still follow a $T^2$ behavior with $A$ = 0.113 $\mu\Omega$ cm/K$^2$. This value doubles the value for YbFe$_2$Zn$_{20}$ (x = 0). The $A$ coefficient keeps increasing with Co concentration: $\sim$0.173 $\mu\Omega$ cm/K$^2$ for x = 0.129 and $\sim$0.197 $\mu\Omega$ cm/K$^2$ for x = 0.170. At x = 0.375, the low-temperature resistivity does not show a decrease and seems to saturate down to 500 mK. As x increases from 0 to 0.375, Yb(Fe$_{1-x}$Co$_x$)$_2$Zn$_{20}$ evolves from a system with a clear Fermi liquid signature in transport to one that does not. 

Fig.~\ref{R}(b) shows resistivity data in the middle of the substitution range (0.375$\leqslant$x$\leqslant$0.875). None of the data shown have a clear signature of a resistivity drop that would be associated with a Kondo coherence. For x = 0.628 (blue star) and 0.719 (green dotted circle), the low-temperature resistivity data show a minor decrease which is followed by an secondary increase at a lower temperature. This is reminiscent of what one would expect from a CEF feature when lowering the temperature depopulates the CEF levels and thus change the degeneracy participating in the Kondo effect\cite{CC72}. However, it is unclear at this point if a coherence might be reached for these substitution levels for temperatures below 500 mK. 

Fig.~\ref{R}(c) presents data close to the pure YbCo$_2$Zn$_{20}$ side (0.875$\leqslant$x$\leqslant$1). Black hollow triangles show the resistivity data for pure YbCo$_2$Zn$_{20}$ which are consistent with the previously reported results\cite{milton}. The resistivity increases with decreasing temperature below $\sim$50 K, indicative of a Kondo effect. The drop of resistivity below 2 K is an indication of Kondo coherence. With a small amount of substitution of Fe for Co, shown by the blue hollow circles (x = 0.991) and green hollow squares (x = 0.986), the temperature at which resistivity starts to drop decreases. With further Fe substitution, the Kondo coherence signature in resistivity could not be observed down to 500 mK as illustrated by purple bar (x = 0.957) and red star data (x = 0.875). Since the coherence temperature of YbCo$_2$Zn$_{20}$ is much lower than that for YbFe$_2$Zn$_{20}$, it takes less Fe substitution to suppress the coherence temperature of YbCo$_2$Zn$_{20}$ to below 500 mK. 

\begin{figure}[tbh!]
\includegraphics[scale = 0.32]{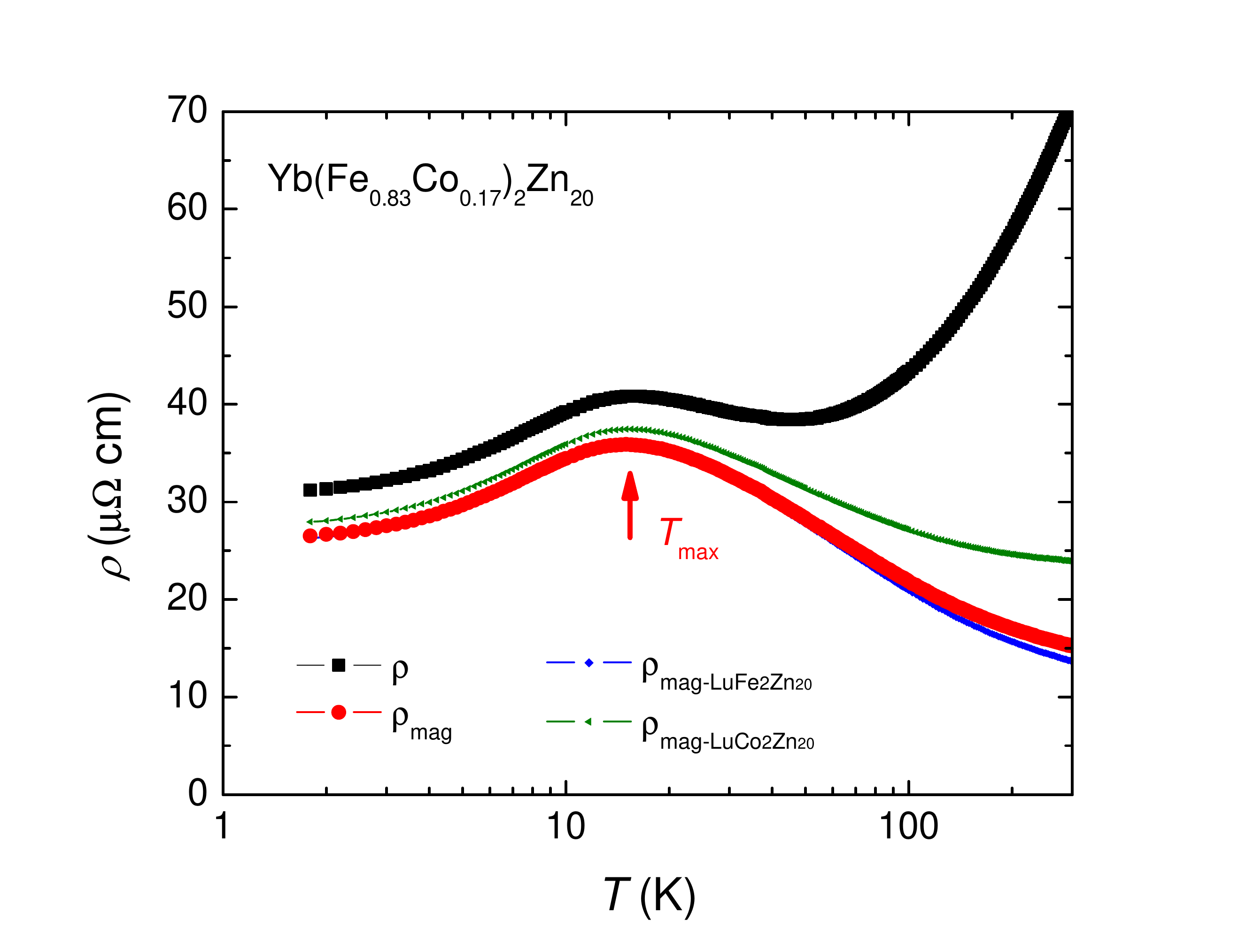}
\caption{(Color online) Temperature-dependent resistivity for Yb(Fe$_{0.83}$Co$_{0.17}$)$_2$Zn$_{20}$ on a semi-log plot. Black squares show the total resistivity. Red circles show the magnetic part of the resistivity. Blue diamonds and green triangles represent magnetic part of resistivity by subtracting the resistivity of pure LuFe$_2$Zn$_{20}$ or LuCo$_2$Zn$_{20}$ (see text). Arrows indicate criteria for determining $T_{max}$ from the $\rho_{mag}$ data.}
\label{criteria}
\end{figure}

Fig.~\ref{criteria} illustrates the criteria that were used to infer characteristic temperatures for the Yb(Fe$_{1-x}$Co$_{x}$)$_{2}$Zn$_{20}$ compounds from the resistivity data. The magnetic part of resistivity was estimated by subtracting a combination of the resistivity for the non-magnetic LuFe$_2$Zn$_{20}$ and LuCo$_2$Zn$_{20}$ with the same Fe/Co ratio. For example, in Fig.~\ref{criteria}, the non-magnetic part of the resistivity data for Yb(Fe$_{0.83}$Co$_{0.17}$)$_2$Zn$_{20}$ were estimated as 0.83$\rho_{LuFe_2Zn_{20}}$ + 0.17$\rho_{LuCo_2Zn_{20}}$. Note, however, such subtraction does not take into account the disorder scattering introduced by Fe/Co substitution. The temperature of the maximum in $\rho_{\text{mag}}$ was tracked as $T_{max}$. For comparison, the magnetic part of the resistivity estimated by subtracting the resistivity of pure LuFe$_2$Zn$_{20}$ (blue diamonds) and LuCo$_{2}$Zn$_{20}$ (green triangles) as well as the resistivity of Yb(Fe$_{0.83}$Co$_{0.17}$)$_2$Zn$_{20}$ (black squares) are also shown in Fig.~\ref{criteria}. The temperatures of the maximum in all these data sets are similar and consistent. In the absence of low-temperature data for non-magnetic subtraction below 1.8 K, the total resistivity data were then used to extract characteristic temperatures. 

\begin{figure*}[tbh!]
\includegraphics[scale = 0.65]{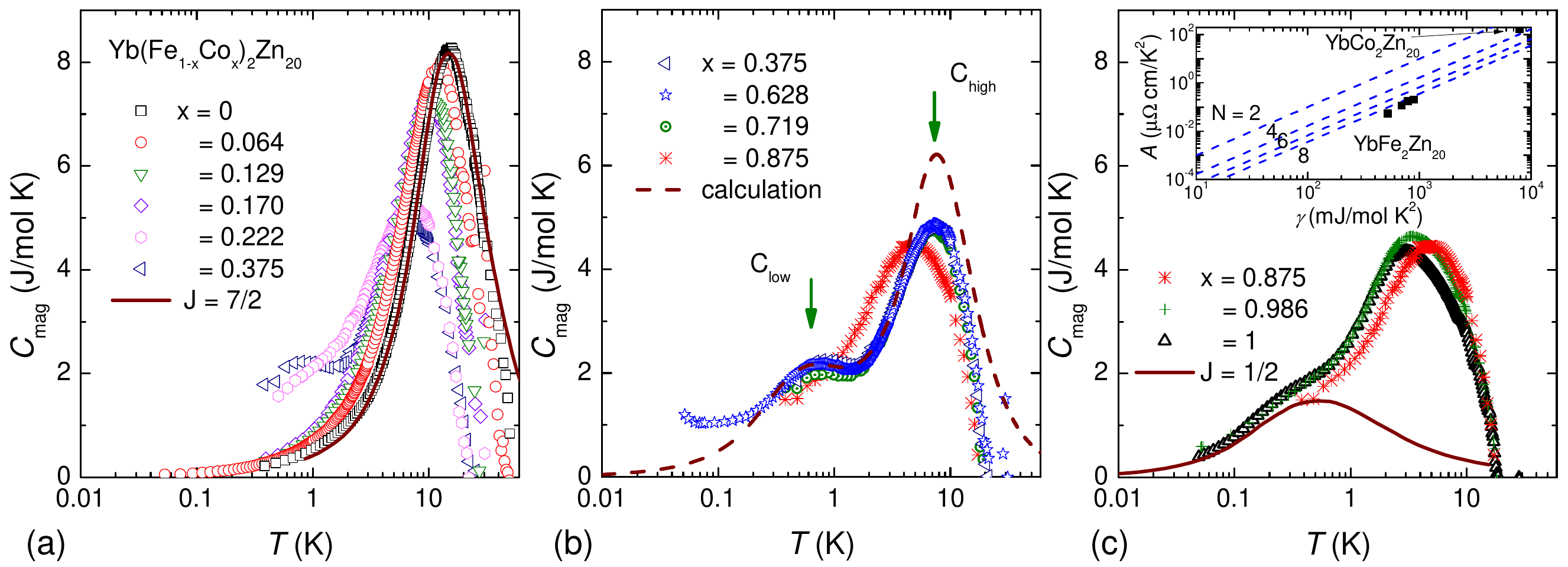}
\caption{(Color online) Temperature-dependent magnetic specific heat of Yb(Fe$_{1-x}$Co$_x$)$_2$Zn$_{20}$ for a various of Co concentrations. Solid lines in (a) and (c) represent magnetic specific heat of the Coqblin-Schrieffer model for J = 7/2 and 1/2\cite{CS69,Rajan83}. The brown dashed line in (b) represents calculated magnetic specific heat based on a model proposed in Ref.~\onlinecite{Romero14} (see text). Green arrows indicate the criteria for determining C$_{\text{high}}$ and C$_{\text{low}}$. Inset in (c) is a generalized Kadowaki-Woods plot with $N$ denoting the degeneracy that is responsible for the Kondo effect\cite{Tsujii05} and data points representing Yb(Fe$_{1-x}$Co$_{x}$)$_{2}$Zn$_{20}$ (x = 0, 0.064, 0.129, 0.170 and 1).}
\label{C}
\end{figure*}

The temperature-dependent magnetic specific heat data for Yb(Fe$_{1-x}$Co$_x$)$_2$Zn$_{20}$ are shown in Fig.~\ref{C}. Similar to the magnetic part of the resistivity, the specific heat of LuCo$_2$Zn$_{20}$ and LuFe$_2$Zn$_{20}$ were used to perform non-magnetic background subtraction. Quantitatively, subtracting LuFe$_2$Zn$_{20}$ or LuCo$_2$Zn$_{20}$ only results in a $\lesssim$2$\%$ difference in the magnetic specific heat value and a $\lesssim$5$\%$ change in characteristic temperature values. Therefore, for all the doped samples, a consistent non-magnetic background specific heat of LuCo$_2$Zn$_{20}$ was subtracted.

The magnetic specific heat of YbFe$_2$Zn$_{20}$ (black squares in Fig.~\ref{C}(a)) can be well explained by a $N$ = 8 Kondo resonance as shown by the brown solid line\cite{Rajan83}. With Co substitution, the Kondo resonance peak moves towards lower temperature accompanied by a decrease in maximum value. At x = 0.064, the electronic specific heat, $\gamma$, increases from 520 mJ/mol K$^2$ for YbFe$_2$Zn$_{20}$\cite{milton} to 690 mJ/mol K$^2$. With more Co substitution, $\gamma$ increases to $\sim$790 mJ/mol K$^2$ for x = 0.129 and $\sim$890 mJ/mol K$^2$ for x = 0.170. Together with the increase in the coefficient of $T^2$ resistivity, the positions of Yb(Fe$_{1-x}$Co$_{x}$)$_{2}$Zn$_{20}$ (x = 0.064, 0.129 and 0.170) on the generalized Kadowaki-Woods plot gradually move toward the YbCo$_2$Zn$_{20}$ side\cite{milton} as shown in the inset of Fig.~\ref{C}(c). As Co substitution increases further, the single peak in YbFe$_2$Zn$_{20}$ gradually evolves to two maxima, as a result of competing energy scale of CEF splitting and the Kondo effect\cite{Desgranges14}. 

In between 0.375$\lesssim$x$\lesssim$0.719, the magnetic specific heat data show very similar behavior. We keep track of the two maxima temperatures as C$_{high}$ (maxima that locates at a higher temperature) and C$_{low}$ (maxima that locates at a lower temperature) [Fig.~\ref{C}(b)]. For x = 0.628, the specific heat was measured down to 50 mK. An upturn at below 100 mK was observed and will be discussed later. With more Co substitution, C$_{high}$ starts to move towards lower temperature.

Fig.~\ref{C}(c) shows magnetic specific heat data close to the pure YbCo$_2$Zn$_{20}$ side. From a generalized Kadowaki-Woods plot, the degeneracy that is responsible for Kondo coherence for YbCo$_2$Zn$_{20}$ is in between 2 and 4. The specific heat can be tentatively understood with a spin 1/2 Kondo resonance with additional contribution from higher temperature Schottky peak-features due to CEF splitting\cite{Takeuchi11,Romero14}. The brown solid line presents the magnetic specific heat due to a spin 1/2 Kondo resonance\cite{Rajan83}. It captures, for the most part, the low-temperature part of the measured data (black triangles), with differences at higher temperatures coming from CEF effects, suggesting a doublet CEF ground state. With addition of Fe, the high-temperature maximum moves towards higher temperature, indicating a small increase of CEF splitting. In the mean time, the low-temperature part of the specific heat sees a slight increase (green squares) at the base temperature. This feature may eventually evolve to an upturn seen for x = 0.628.

\section{Discussion}

\begin{figure}[tbh!]
\includegraphics[scale = 0.35]{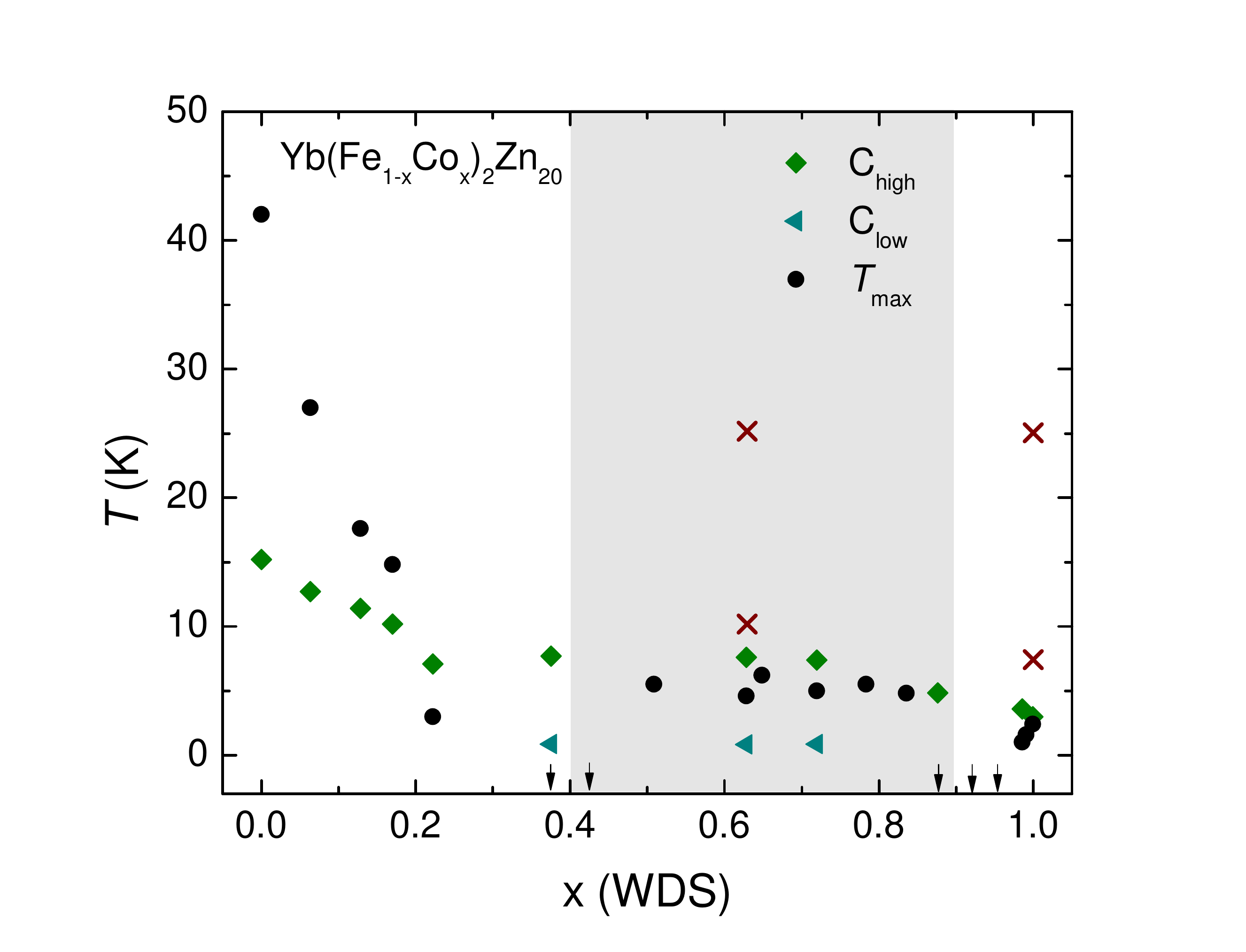}
\caption{(Color online) Characteristic temperatures as a function of Co concentration in Yb(Fe$_{1-x}$Co$_x$)$_2$Zn$_{20}$. Green solid diamond and cyan triangle represent specific heat maxima temperatures at high temperature and low temperature respectively. Black circles represent $T_{max}$ values extracted from resistivity data as illustrated in Fig.~\ref{criteria}. Arrows at the bottom indicate samples that were measured resistively but showed no $T_{max}$ down to 2 K (x = 0.426, 0.921) or 500 mK (x = 0.375, 0.875, 0.957). Brown crosses indicate the estimated energy of excited CEF levels for x = 0.628 and 1. The grey shaded area separates the phase diagram into three regions as described in the text.}
\label{summ}
\end{figure}

Summarizing resistive and specific heat features presented above, we can plot characteristic temperatures of Yb(Fe$_{1-x}$Co$_x$)$_2$Zn$_{20}$ as a function of Co concentration, x. In Fig.~\ref{summ}, the green diamonds and cyan triangles represent characteristic temperatures inferred from specific heat data and black circles were inferred from resistivity measurements. At a gross level, the phase diagram can be divided into three regions. Two regions are on the pure YbFe$_{2}$Zn$_{20}$ or YbCo$_{2}$Zn$_{20}$ sides where the original Kondo lattice characteristic temperatures gradually evolve with Fe/Co substitution. The third region is in the middle where all feature temperatures are relatively similar and do not significantly change or evolve with x.

We first look at the two regions close to the pure YbFe$_2$Zn$_{20}$ and YbCo$_{2}$Zn$_{20}$ sides. The crossover from a high-temperature, single ion regime to a low-temperature coherent regime can usually be seen and inferred from temperature-dependent resistivity measurements\cite{Cox1988, Bauer91}. In the coherent regime, the resistivity drops at low-temperature, becoming a heavy Fermi-liquid state. Both YbFe$_{2}$Zn$_{20}$ and YbCo$_{2}$Zn$_{20}$ have a heavy Fermi liquid ground state down to 50 mK as evidenced by linear specific heat and a $T^{2}$ dependence in resistivity\cite{milton,Canfield08}. On the YbFe$_{2}$Zn$_{20}$ side, Co substitution suppresses $T_{max}$ at roughly 1.8 K/$\%$Co. At x = 0.222, the $T_{max}$ is suppressed to 3 K. With more Co substitution, $T_{max}$ was suppressed below the base temperature of measurements. A similar situation happens on the YbCo$_{2}$Zn$_{20}$ side. $T_{max}$ is suppressed upon Fe substitution at roughly 1 K/$\%$Fe. Given a much smaller, initial $T_{max}$ value to start with, the trackable features quickly disappears to below 50 mK.

\begin{figure}[tbh!]
\includegraphics[scale = 0.35]{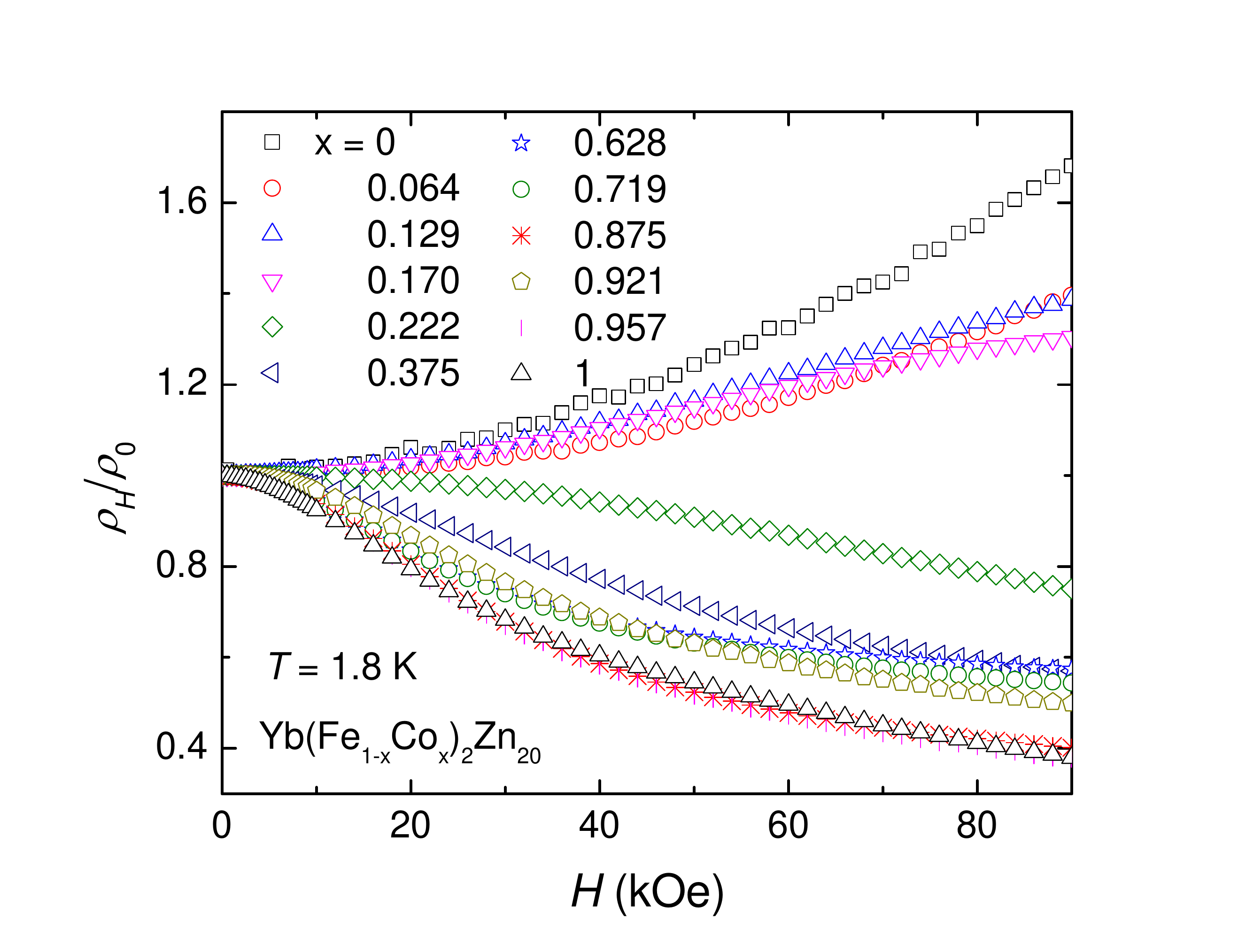}
\caption{(Color online) Magnetoresistance of Yb(Fe$_{1-x}$Co$_{x}$)$_{2}$Zn$_{20}$ measured at 1.8 K up to 90 kOe.}
\label{RH}
\end{figure}

Magnetoresistance is sometimes used to probe the Kondo state at low temperature as well\cite{Aronson89}. The single ion Kondo regime has been theoretically calculated to show a negative magnetoresistance\cite{Schlottmann83}. In the coherent regime, the compound is essentially a heavy Pauli-paramagnetic metal. Magnetoresistance therefore is most commonly positive. The magnetoresistance for YbFe$_{2}$Zn$_{20}$ is positive at 1.8 K, which is consistent with a coherent state (Fig.~\ref{RH}). For YbCo$_{2}$Zn$_{20}$, prior to the metamagnetic transition, the magnetoresistance is also positive below 0.1 K whereas shows negative magnetoresistance above 3 K\cite{Saiga09,Honda14}. 

Magnetoresistance data are shown for various Co concentrations at 1.8 K in Fig.~\ref{RH}. Positive magnetoresistances for Co concentrations x $\lesssim$ 0.170 suggest a coherence at 1.8 K. Other members (x$\geqslant$0.222) show negative magnetoresistances that suggest a single ion state. Assuming that the $T_{max}$ is commensurate with coherence temperature, together with the magnetoresistance data, the coherence is suppressed on both ends of the phase diagram (Fig.~\ref{summ}). In the middle of the substitution range, however, no clear indication of coherence could be determined at 1.8 K. %Note: The magnetoresistance have the most negative slope when H is close to high field Kondo energy scale.%

Specific heat features close to the pure YbFe$_2$Zn$_{20}$ and YbCo$_2$Zn$_{20}$ sides evolve gradually with chemical substitution. On the YbFe$_{2}$Zn$_{20}$ side, with Co substitution, C$_{high}$ moves to lower temperature, indicating a lowering of Kondo temperature\cite{Rajan83}. On the pure YbCo$_2$Zn$_{20}$ side, where the Kondo temperature is smaller than the CEF splitting, with Fe substitution, C$_{high}$ moves towards higher temperature, indicating a slight increase of CEF splitting.

The Kondo coherence feature in specific heat is much more subtle and under debate. Various models have proposed a decrease of density of states at the Fermi energy due the formation of coherence\cite{Martin82,Grewe84,Lacroix86,Kaga88}. Experimentally, this decrease of density of states was used to explain the low-temperature drop in $C/T$ (or $\gamma$) in different systems, for example, in CeAl$_3$\cite{Brodale86, Bredl84}, CuCu$_6$\cite{Fujita85}, CeNi$_2$Ge$_2$\cite{Knopp88} and CeCu$_2$Si$_2$\cite{Bredl84}. However, a feature associated with coherence in specific heat is not always so apparent and is commonly missing, for example in YbNi$_2$B$_2$C\cite{Yatskar96b}, YbCuAl\cite{Schlottmann1989}, CeCoGe$_2$\cite{Mun04}, YbAgCu$_{4}$\cite{Rossel87, Besnus90} and CeNi$_9$Si$_4$\cite{Michor03}. A single ion model based on a Coqblin-Schrieffer analysis\cite{CS69, Rajan83} seems to describe these Kondo lattice systems very well, even in the coherent regime. For Yb$TM_2$Zn$_{20}$, despite the fact that coherence emerges in resistivity at low-temperature, specific heat can be well captured by single ion Kondo effect\cite{milton}.

Close to pure YbFe$_2$Zn$_{20}$ and YbCo$_2$Zn$_{20}$, it is shown above that upon Fe/Co substitution, the coherence temperatures on both sides are suppressed. The suppression of coherence is commonly achieved by substituting the moment bearing ions with non-moment bearing ions\cite{Stewart01}. In the case of Yb(Fe$_{1-x}$Co$_{x}$)$_{2}$Zn$_{20}$, moment bearing Yb-sites are always fully occupied and ordered. Substituting Fe/Co, however, will inevitably bring in chemical disorder as well as a change in band filling. As a consequence of chemical disorder, the decrease in mean free path of the conduction electrons, as can be represented by the increase of residual resistivity, may lead to a suppression of Kondo coherence temperature\cite{Paglione07}. This is sometimes also seen in systems with artificially created defects with irradiation\cite{Adrian89}. Theoretically, it is proposed that disorder on the moment-bearing site (f-site) affects Kondo coherence more efficiently than disorder on the conduction-electron sites. To produce the same suppression of Kondo coherence temperature, energetically, f-sites disorder need to be equivalent to $T_{K}$ whereas conduction-electron sites need to be comparable to the band width of the conduction band, which is more difficult\cite{ZT86}. However, with a change of band filling due to Fe/Co substitution, disorder in the conduction electron sites could affect Kondo coherence in these Yb(Fe$_{1-x}$Co$_{x}$)$_{2}$Zn$_{20}$ compounds.

The change in band filling due to Fe/Co substitution, on one hand, could be part of the disorder in conduction electrons in aforementioned theoretical model. On the other hand, it may lead to a change in the ratio of Kondo coherence temperature over Kondo temperature, $T_{coh}$/$T_{K}$\cite{Pruschke00}. At half-filling, it is proposed that $T_{coh}$ is larger than $T_{K}$. Away from half-filling, $T_{coh}$ drops quickly below $T_{K}$\cite{Pruschke00}. It is possible that with changing band filling, $T_{coh}$ decreases from both the YbFe$_2$Zn$_{20}$ and YbCo$_2$Zn$_{20}$ sides, which leaves a clear single ion Kondo effect with features associated with CEF population whereas showing no coherence or a very low coherence temperature. Such an extended range of low $T_{coh}$ is observed in the middle region of the phase diagram. A change in band filling may be inferred from the density of states at Fermi energy. It worth noting that the density of states for YFe$_2$Zn$_{20}$ experiences a quick drop upon adding Co and stays relatively constant above 20$\%$ Co substitution\cite{shuang2007}. This is very similar to the suppression of $T_{max}$ close to the YbFe$_2$Zn$_{20}$ side. Close to the YCo$_2$Zn$_{20}$ side, though, there are not enough data density to make similar comparison to YbCo$_2$Zn$_{20}$.

%One of the experimental techniques that could measure density of states is Hall effect. Hall coefficient of YbCo$_{2}$Zn$_{20}$ was previously reported down to 50 mK\cite{Tomooka12}. The Hall coefficient shows a broad maximum at around 2 K, a temperature that is very similar with $\rho_{max}$ in resistivity. It also shows an upturn below 400 mK. The overall shape of the Hall coefficient can be qualitatively pictured by skew scattering\cite{Fert87}. However, it is difficult to quantitatively extract the ordinary Hall effect that allows for an estimation of density of states. It might be possible that scattering mechanisms are needed to better fit the observed data\cite{Kontani94,Yang13}. It worth further study to look at the band filling across the substitution series.

We now move onto the middle region of the phase diagram shown in Fig.~\ref{summ}. Whereas there are clear changes of characteristic temperatures as the sample departs from perfect chemical order, in the middle region, for 0.4$\lesssim$x$\lesssim$0.9, characteristic temperatures stay fairly constant. It worth noting though, as shown in Fig.~\ref{WDS}, the variation in concentration across a sample in this middle region may make the data blurred and offer only more qualitative information rather than quantitative.

In a single ion Kondo picture, different local moment degeneracy can give rise to unique features in thermodynamic and transport properties\cite{CS69,CC72,BC76}. In the presence of crystalline electric field, features can be observed in temperature-dependent transport measurements, exhibited as broad maxima in resistivity, for example in CeAl$_{2}$\cite{NF72}, CeCu$_{2}$Si$_{2}$\cite{Schneider83}, CePdIn\cite{Fujii92}, CeZn$_{11}$\cite{Taufour13}. This was observed for several members of Yb(Fe$_{1-x}$Co$_{x}$)$_{2}$Zn$_{20}$ series, in the middle substitution range, for example, x = 0.628 and 0.719. However, the CEF feature is not as clear as systems mentioned above. If the CEF levels are not well separated, given a certain combination of density of states and Kondo coupling, these broad maxima can be hard to observe\cite{CC72}. In the case of Yb$TM_2$Zn$_{20}$, the CEF splitting is indeed small and may cause the CEF feature to be difficult to observe. Impurity scattering can also lead to a smear of Kondo CEF features. In YbNi$_{2}$B$_{2}$C, the improved sample quality after annealing dramatically changed the temperature-dependent transport properties. This can be attributed to a distribution of Kondo temperature in a non-ideal lattice with local defects and strain\cite{Avila04}. In Yb(Fe$_{1-x}$Co$_{x}$)$_{2}$Zn$_{20}$, the substitution variation in the sample may cause such disorders even though transition metal is not in the direct neighborhood of the Yb ions.

To understand what happens in the middle of the substitution range in Yb(Fe$_{1-x}$Co$_{x}$)$_{2}$Zn$_{20}$, more insights can be obtained from the specific heat data. In the presence of crystal fields, the temperature-dependence of magnetic specific heat in Kondo systems is complicated. An arbitrary CEF splitting has only been recently studied numerically for Ce-based compounds\cite{Desgranges14}. For Yb, more degeneracy, and levels, are involved. A quantitative interpretation of the temperature-dependent specific heat can be approached by combining a resonance model solution together with a, CEF, Schottky-like contribution\cite{Romero14}. In a cubic symmetry, Yb will be split into two doublets and a quartet. Assuming the quartet is at higher temperature, the temperature-dependent specific heat can be written as\cite{Romero14}:

\begin{equation}
C_{mag} = C_{2d} - \frac{1}{k_{B}T^{2}}\bigg[\frac{\Delta^{2}e^{-\Delta/k_{B}T}}{(1+e^{-\Delta/k_{B}T})^{2}}\bigg] + C_{s}
\end{equation}

in which,

\begin{multline}
C_{2d} = -\frac{k_{B}}{2(\pi k_{B} T)^{2}} Re \sum_{j=0}^{1}\bigg\{(\Gamma_{j}+i\Delta_{j})^2\bigg[4\psi'(\frac{\Gamma_{j}+i\Delta_{j}}{\pi k_{B}T}) \\
-\psi'(\frac{\Gamma_{j}+i\Delta_{j}}{2\pi k_{B}T})\bigg]\bigg\} + \frac{\Gamma_{0}+\Gamma_{1}}{\pi T}
\end{multline}

\begin{multline}
C_s = \frac{1}{k_B T^2}[(\Delta_1)^2 e^{-\Delta_1/k_B T} + 2(\Delta_2)^2 e^{-\Delta_2/k_B T}\\
+2(\Delta_2-\Delta_1)^2e^{-(\Delta_1+\Delta_2)/k_B T}](1+e^{-\Delta_1/k_B T} + 2e^{-\Delta_2/k_B T})^{-2}
\end{multline}

Here, C$_{2d}$ is the Kondo resonance contribution from the lower lying two doublets. $\Gamma_{j}$ represents the half-width at half-maximum of the spectral density for each crystal field level. $\psi'$ is the derivative of the digamma function and C$_{s}$ is th Schottky expression for a three level system. $\Delta_{j}$ represents the excited CEF energies. The second term in Eq.(1) accounts for the double counting of the Schottky contribution. $\Delta$ equals to $\Delta_{1}$ ignoring complications of the ground state doublet\cite{Romero14}. $\Delta_0$ for the ground state doublet is introduced to account for the resonance that is displaced away from the Fermi energy\cite{Romero14}. 

In Fig.~\ref{C}(b), data for Yb(Fe$_{0.372}$Co$_{0.628}$)$_{2}$Zn$_{20}$ could be best fitted by the brown dashed line with parameters: $\Delta_0$ = 1 K, $\Delta_1$  = 10 K, $\Delta_2$ = 25 K, $\Gamma_0$ = 0.87 K, $\Gamma_1$ = 4 K. The low-temperature rise is omitted in the fit and will be discussed below. C$_{low}$ mainly comes from the Kondo effect for the ground state doublet and C$_{high}$ largely comes from Schottky contribution due to CEF population. Therefore, the temperature of C$_{high}$ also reflects an upper limit of the total CEF splitting.

The calculated value qualitatively agrees with experimental data. Higher calculated values around 10 K might due to an error caused by non-magnetic background subtraction as illustrated by a negative value of C$_{mag}$ for $T >$ 20 K. It could also due to a lack of bandwidth information for the highest lying quartet in the theoretical model\cite{Romero14}. In general, the best fit indicates that the two CEF levels are at around 10 K and 25 K, which is similar to what had been proposed for YbCo$_{2}$Zn$_{20}$\cite{Takeuchi11,Romero14}. Thus at least to this level of Co substitution, as shown in Fig.~\ref{summ} for Co-rich side of the phase diagram, the CEF splitting does not change significantly from the pure YbCo$_2$Zn$_{20}$. In addition, since the temperature of C$_{high}$ sets an upper limit of the total CEF splitting, the total CEF splitting does not change much across the whole substitution range and may have increased slightly when approaching the YbFe$_2$Zn$_{20}$ side. In contrast, the CEF effect is more apparent for YbCo$_2$Zn$_{20}$. The difference in Kondo physics between YbFe$_2$Zn$_{20}$ and YbCo$_2$Zn$_{20}$ most likely originate from the difference in density of states as well as Kondo coupling strength.

Going back to the generalized Kadowaki-Woods plot shown in Fig.~\ref{C}(c), upon Co substitution up to x = 0.170, the degeneracy that is responsible for the Kondo effect is still very close to the full degeneracy: $N$ = 8. However, the temperatures at which Kondo coherence is developed have become comparable or even smaller than the estimated CEF splitting. This trend can be clearly represented by $T_{max}$ shown in Fig.~\ref{summ}. It may be possible that, for x = 0.170 as an example, the Kondo temperature is still at a relatively higher temperature than the CEF splitting, which quenches the local moment with a degeneracy close to 8\cite{Desgranges14}. On the other hand, Kondo coherence, as tracked by $T_{max}$, does not significantly affect the way magnetic entropy is removed. With an even lower Kondo energy scale, the CEF feature would have enough room to be more apparent, like in the case for x = 0.222. 

Finally, as for the low-temperature rise in the specific heat of Yb(Fe$_{0.372}$Co$_{0.628}$)$_{2}$Zn$_{20}$, since there is no magnetic ordering or applied magnetic field, there should not be a nuclear Schottky anomaly due to Zeeman splitting of nuclear levels in specific heat measurement. However, the low-temperature upturn in the specific heat data observed below 100 mK could still come from a nuclear quadrupolar splitting of the $^{173}$Yb nuclear moment\cite{Steppke10}. Such quadrupole splitting for Yb in a cubic point symmetry may arise from the electric field gradient caused by chemical substitution induced distortion. Even though transition metal is not the first, nor the second nearest neighbors of Yb, as substitution increases, the distortion to the original cubic symmetry increase. This is consistent with our observation that the low-temperature rise only emerges with chemical substitution and becomes more pronounced in the middle of the doping range. An alternative scenario is that the upturn is of a Kondo origin. However, that requires a decreasing CEF split energies upon substitution and a very small first excited CEF energy which was not observed in specific heat data. Instead, the C$_{low}$ feature stays unchanged for the majority of the substitution range which otherwise should also evolve and split.

We would also like to mention a similar doping series: CeNi$_{9}X_{4}$ ($X$ = Si, Ge)\cite{Gold12}. Changing from CeNi$_{9}$Ge$_{4}$ to CeNi$_{9}$Si$_{4}$, the Kondo temperature increases from $\sim$4 K to $\sim$70 K with ground state CEF degeneracy changing from 4 for $X$ = Ge to 6 for $X$ = Si\cite{Gold12}. Upon doping, the coherence temperature drops quickly from both sides. In the middle of the doping, specific heat shows continuous evolution from high $\gamma$, Ge side, to the low $\gamma$, Si side. Understanding the suppression of Kondo coherence in these two systems might offer useful insights into the formation of Kondo coherence.

\section{Conclusions}

In conclusion, we studied the evolution of Kondo effect in Yb(Fe$_{1-x}$Co$_x$)$_2$Zn$_{20}$ via resistivity and specific heat measurements. With Co substitution, the Kondo coherence temperature of YbFe$_2$Zn$_{20}$ decreases gradually with emerging features in specific heat that can be associated with CEF effect. On the YbCo$_{2}$Zn$_{20}$ side, the coherence temperature is also suppressed at the beginning of Fe substitution. In between, 0.4$\lesssim$x$\lesssim$0.9, CEF features can be observed in both resistivity and specific heat data whereas showing no clear feature of coherence down to 500 mK. However, only qualitative information can be obtained in this middle region due to a large substitution level variation. Comparing all the experimental results, the CEF splitting stays roughly unchanged across the substitution series. The ground state of the compound evolves from a $N$ = 8 coherent state for YbFe$_2$Zn$_{20}$ to a $N$ = 2 coherent state in YbCo$_2$Zn$_{20}$. More measurements are needed to reveal the mechanism behind the suppression of coherence on the YbFe$_2$Zn$_{20}$ and YbCo$_2$Zn$_{20}$ side.

\section*{Acknowledgements}
We would like to thank U. (G. D. M.) Kaluarachichi, K. Cho, G. Drachuk, B. Song, Y. Furukawa, R. Flint for useful discussions, W. Straszheim for WDS analysis. This work was supported by the U.S. Department of Energy (DOE), Office of Science, Basic Energy Sciences, Materials Science and Engineering Division. The research was performed at the Ames Laboratory, which is operated for the U.S. DOE by Iowa State University under contract NO. DE-AC02-07CH11358.

\bibliographystyle{apsrev4-1}
%\bibliography{references}
%

\end{document}